\documentclass[preprint,showpacs,preprintnumbers,amsmath,amssymb]{revtex4}

\usepackage{graphicx}
\usepackage{dcolumn}
\usepackage{bm}

\begin{document}

\title{\large  Magnetic proximity effect of a topological insulator and a ferromagnet
in thin film bilayers of $\rm Bi_{0.5}Sb_{1.5}Te_3$ and $\rm SrRuO_3$ }

\author{Gad Koren}
\email{gkoren@physics.technion.ac.il} \affiliation{Physics
Department, Technion - Israel Institute of Technology Haifa,
32000, ISRAEL} \homepage{http://physics.technion.ac.il/~gkoren}

\date{\today}
\def\bfig {\begin{figure}[tbhp] \centering}
\def\efig {\end{figure}}

\normalsize \baselineskip=8mm  \vspace{15mm}



\begin{abstract}

Magnetic proximity effect of a topological insulator in contact with a ferromagnet is reported in thin film bilayers of 15 nm thick $\rm Bi_{0.5}Sb_{1.5}Te_3$ on either 15 or 40 nm thick $\rm SrRuO_3$ on (100) $\rm SrTiO_3$ wafers. $\rm SrRuO_3$ is an itinerant ferromagnet which has long been considered weak, thus any observation of a significant magnetic proximity effect in the present system should help elucidate the mechanism of this magnetism and might be utilized in device applications. Magneto transport results of the bilayers were compared with those of reference films of 15 nm $\rm Bi_{0.5}Sb_{1.5}Te_3$ and 15 or 40 nm $\rm SrRuO_3$. Comparison of the temperature coefficient of resistance [(1/R)$\times$dR/dT which is qualitatively proportional to the magnetization] of the bilayer and reference ferromagnetic film normalized above $\rm T_c$, shows a clear suppression in the bilayer by about 50\% just below $\rm T_c$, indicating a weaker proximity magnetization in the bilayer. Resistance hysteresis loops versus field at 1.85$\pm$0.05 K in the bilayer and reference films show a clear magnetic proximity effect, where the peak resistance of the bilayer at the coercive field shifts to lower fields by $\sim$30\% compared to a hypothetical bilayer of two resistors connected in parallel with no interaction between the layers. Narrowing of the coercive peaks of the bilayers as compared to those of the reference ferromagnetic films by 25-35\% was also observed, which represents another signature of the magnetic proximity effect.  \\


\end{abstract}

\maketitle

\section{Introduction}
\normalsize \baselineskip=6mm  \vspace{6mm}

Interactions between topological insulators (TI) and magnetic materials, either ferromagnets (FM) or antiferromagnets (AFM), are interesting due to the basic physics questions they raise as well as their possible potential application in spintronics and quantum computing \cite{KaneRMP,Moore,Kitaev,Vorobnik,YangAndo,LiMoodera,LiMoodera2,WeiMoodera,YangKapitulnik,HesejdalChen,Weng}.
These interactions which basically originate in magnetic proximity effects (MPE), can induce long-range ferromagnetism in magnetically doped TI at room temperature \cite{Vorobnik,Katmis}. They can lead to spin or magnetization sensitive switching phenomena \cite{YangAndo,Qing,Yabin}, to quantum anomalous Hall effect (QHE without external magnetic field) \cite{LiMoodera,WeiMoodera,Weng,Zilong,Liu,Chang, Kandala,Kou}, to weak-localization-like magnetoresistance effects \cite{YangKapitulnik} and so on. Some of the data is hard to reconcile with conventional MPE, such as the observation of Wei et al. \cite{WeiMoodera} showing that in their TI-FM bilayer with 1 nm thick FM layer the magnetic moment was found to be larger by 60\% than that of the FM magnetic ions. A conventional proximity effect can never induce stronger ferromagnetism in TI-FM bilayers just as it can't induce stronger superconductivity in TI-S bilayers (where S is a superconductor) \cite{Koren}. Therefore, it seems that unconventional MPE can be involved in TI-FM interfaces. MPE was observed even in S-FM heterostructures where evidence for induced magnetization in the superconductor was found by scanning electron spectroscopy \cite{Asulin}. Another issue is the sign of the induced magnetoresistance which can be positive or negative depending on the system involved \cite{WeiMoodera,YangKapitulnik,OldMoodera}. In the present study, TI-FM bilayers of the topological insulator $\rm Bi_{0.5}Sb_{1.5}Te_3$ (BST) and the supposedly weak itinerant ferromagnet $\rm SrRuO_3$ (SRO) with a ferromagnetic transition temperature $\rm T_c=150$ K are investigated, and a surprisingly strong conventional MPE with positive magnetoresistance is found.\\

\section{Preparation and basic properties of the films\\
and bilayers }
\normalsize \baselineskip=6mm  \vspace{6mm}

SRO thin films on (100) $\rm SrTiO_3$ (STO) were chosen for this study as the ferromagnetic layers since these films are very well characterized in the literature \cite{Marshall,Klein}. $\rm (Bi_xSb_{1-x})_2Te_3$ thin films were chosen as the topological insulator layers for the ability to tune their Fermi energy $\rm E_F$ to the Dirac point by changing the Bi doping x \cite{Zhang}. Both were prepared by laser ablation deposition  in different vacuum chambers using the third harmonic of a Nd-YAG laser. Deposition was made on (100) STO wafers of $10\times 10$ mm$^2$ area. The SRO films were deposited using a stoichiometric ceramic target of SRO under 70-90 mTorr of $\rm O_2$ gas flow at 965 $^0$C heater block temperature (about 800 $^0$C on the surface of the wafer). The BST layers were deposited using a $\rm Bi_{0.2}Sb_{1.8}Te_3$ target under vacuum and at 320 $^0$C heater block temperature (about 250 $^0$C on the surface of the wafer). Electron dispersive spectroscopy (EDS) measurements \cite{supplement} showed that the resulting films were about 2.5 times richer in Bi than the target used yielding $\rm Bi_{0.5}Sb_{1.5}Te_3$. Thus laser deposition is not as good as molecular beam epitaxy (MBE) for controlling the Bi and Sb content in these films \cite{Zhang}. The laser was operated at a pulse rate of 3.33 Hz, with fluence of $\rm \sim 1.5\, J/cm^2$ on the target for the deposition of the SRO films, and lower fluence of $\rm \sim 0.6\, J/cm^2$ for the deposition of the BST layers. X-ray diffraction measurements of our typical 15 nm thick SRO reference film, and 15 nm thick BST on 15 nm thick SRO bilayer, showed that the SRO film grew epitaxially on the (100) STO wafer, while the BST cap layer grew in the hexagonal phase with preferential c-axis orientation normal to the wafer with c=3.04 nm \cite{supplement}. This yields 5 unit cells in a 15 nm thick BST layer, leading to strong interactions between the top and bottom surface currents and basically resulting in a 2D film. Atomic force microscopy (AFM) images of the surface morphology of the 30 nm thick bilayer showed good crystallization and $\sim$3-4 nm rms roughness \cite{supplement}.    \\

Bilayers (BLs) with thicker SRO layers were also prepared and characterized. These comprised of 15 nm thick BST film on 40 nm thick SRO layer on (100) STO wafers. Here we also added patterning of the SRO films and BST-SRO bilayers into a 30 cm long meander-line with 18 $\mu$m line-width and 2 $\mu$m line-spacing. Patterning was done under 13 mA/cm$^2$ Ar-ion milling at -180 $^0$C to minimize damage to the samples. This allowed us to increase the measurements sensitivity by having a better signal to noise ratio, but at the cost of having a more dominant SRO signal and less sensitivity to the TI layer properties. Nevertheless, the basic MPE property in the BL of having narrower magnetoresistance peaks at the coercive fields as compared to those of the reference SRO film, could already be seen also in this case. The transport measurements were done by the use of an array of 40 gold coated spring loaded spherical tips for the 4-probe dc measurements on ten different locations on the wafer. All measurements on the un-patterned 15 nm BST on 15 nm SRO bilayer and the 15 nm SRO reference film were made without additional contact pads (to keep surface cleanliness in between the deposition steps), while measurements on the meander-lines were carried out using silver paste contact pads.   \\

\section{Results and discussion}

\subsection{15 nm BST on 15 nm SRO bilayers and  15 nm SRO reference films}

Fig. 1 shows the resistance versus temperature results of a 30 nm thick bilayer of 15 nm BST on 15 nm SRO together with the results on the 15 nm SRO reference film normalized to that of the bilayer at room temperature. One can easily see the bending down of both curves at the  ferromagnetic transition at 150 K. These R versus T curves under ZFC are quite close above $\rm T_c$ but becoming farther apart on cooling down below it. To demonstrate this more clearly, we plot here also the normalized ratio of the two curves R(BL ZFC)/[0.22$\times$R(SRO ZFC)], which clearly shows an almost constant behavior versus T above $\rm T_c$ and increasing values toward lower temperatures just below it. This is a clear signature of a magnetic proximity effect (MPE) resulting from a suppressed rate of resistance decrease with decreasing temperature of the BL than that of the SRO film. Furthermore, should there be no interaction between the layers in the bilayer, adding the BST resistor in parallel to the SRO one, should decrease the BL resistance and not increase it as is actually observed using the normalized SRO data in Fig. 1. Thus the existence of MPE in the BL follows. Also shown in this figure is the R(T) under 1 T field cooling (FC) of the bilayer. The ZFC and the FC curves are quite indistinguishable with a very small magnetoresistance, which does not show any systematic behavior versus T, probably due to noise in the measurements.\\

\begin{figure} \hspace{-20mm}
\includegraphics[height=6cm,width=8cm]{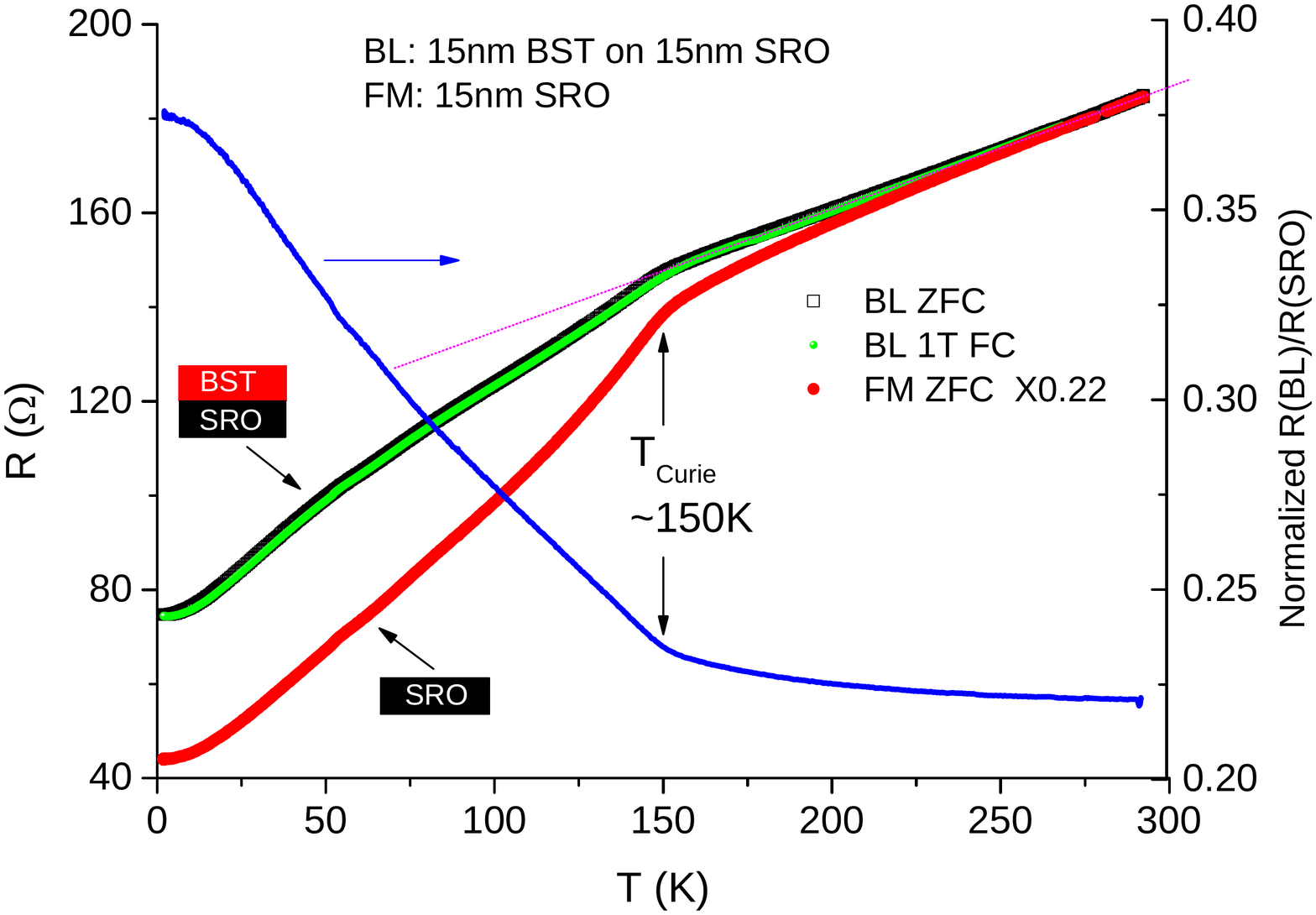}
\hspace{-20mm} \caption{\label{fig:epsart} (Color online) Normalized resistance versus temperature of a 15 nm thick SRO film under zero field cooling, together with the corresponding results of the bilayer obtained by the deposition of a 15 nm thick BST layer on it. Also shown are R versus T curve of the bilayer under 1 T field cooling, and the normalized ratio R(BL ZFC)/[0.22$\times$R(SRO ZFC)] versus T.     }
\end{figure}

\begin{figure} \hspace{-20mm}
\includegraphics[height=6cm,width=8cm]{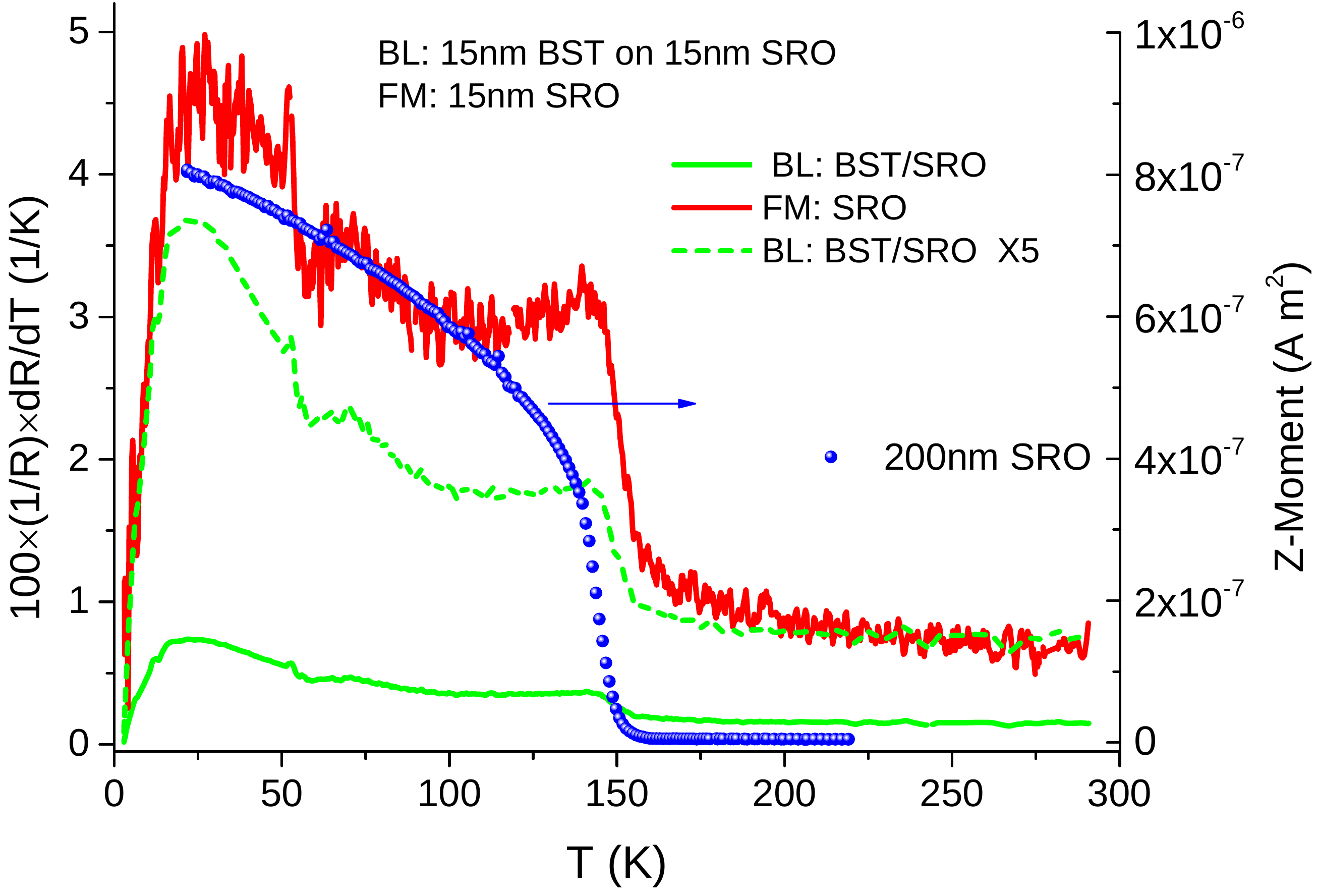}
\hspace{-20mm} \caption{\label{fig:epsart} (Color online) Temperature coefficients of resistance (TCR) in percents per K [100$\times$(1/R)$\times$dR/dT] are plotted versus T for the 15 nm BST on 15 nm SRO bilayer and the reference 15 nm SRO film. Also shown is the Z-moment of a 200 nm thick SRO film as measured by a SQUID magnetometer.  }
\end{figure}

Fig. 2 presents the temperature coefficients of resistance (TCR) in percents per K [100$\times$(1/R)$\times$dR/dT] versus temperature under zero field of the 15 nm BST on 15 nm SRO bilayer and of the 15 nm SRO reference film. These coefficient are useful indicators for the magnetic properties of the bilayer and film since they are qualitatively similar to the magnetization of the samples, as can be seen by the magnetic moment of a 200 nm thick SRO film on (100) STO versus T which is also plotted in this figure. This moment was measured by a SQUID magnetometer, and the reason for using such a thick film here is that the signal from the thinner 15 nm SRO was too noisy. The excess noise in the TCR curve of the 15 nm SRO film as compared to that of the bilayer is due to the noisy contacts on this very thin film as no contact pads were used in this measurement. This preserved a clean surface of this film which was then used for the deposition of the cap 15 nm BST layer on it to produce the bilayer. One can see that the sharp rise of the TCR and Z-moment at $T_c$=150 K is basically common to all curves. Also, above this temperature the curves are almost constants with a small fluctuation regime above $T_c$. Below 140 K, these curves although rising with decreasing temperature down to about 20 K, are not proportional to one another and there is additional information in the TCR curves whose origin is unclear at the present time. Below 20 K, the resistance of the BL and SRO film becomes quite constant (see Fig. 1), thus the TCR obviously goes to zero while the magnetization doesn't. We normalized the TCR curve of the bilayer to that of the SRO reference film in the paramagnetic regime above $T_c$ in the range of 200-290 K in order to allow comparison between the two. We see that the TCR of the bilayer is suppressed by about 50\% as compared to that of the reference film and this is a clear indication for the presence of a magnetic proximity effect in the bilayer. As no enhancement of the TCR in the bilayer is found as in \cite{WeiMoodera}, we conclude that the present results support the notion of a conventional MPE here.\\

\begin{figure} \hspace{-20mm}
\includegraphics[height=6cm,width=8cm]{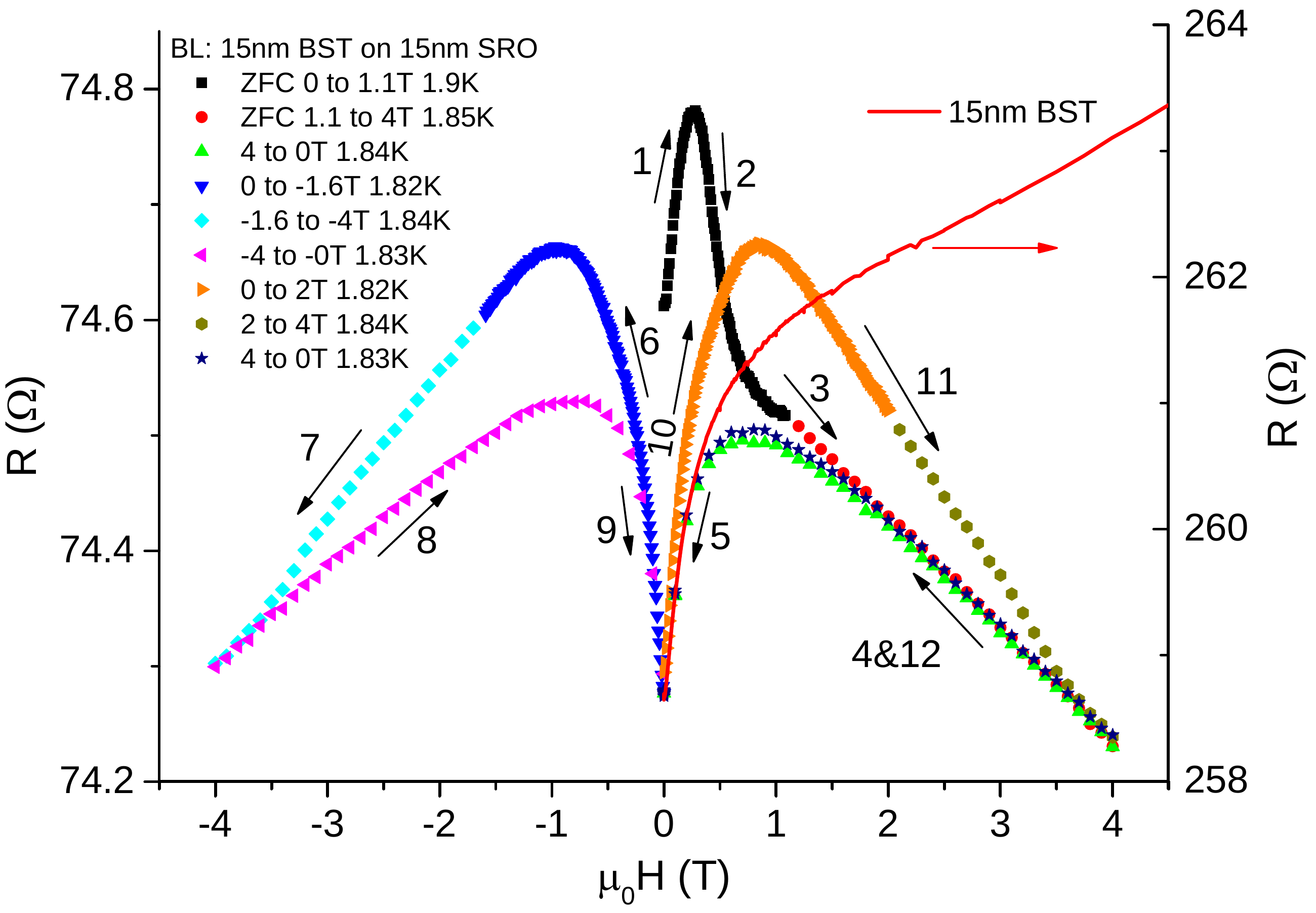}
\hspace{-20mm} \caption{\label{fig:epsart} (Color online) Resistance hysteresis loops versus field normal to the wafer at 1.85$\pm$0.05 K of the 15 nm BST on 15 nm SRO bilayer, together with the non-hysteretic magnetoresistance of a 15 nm thick BST film (red curve). The black arrows and numbers follow the magnetic history of the measurements.      }
\end{figure}

Next we turn to resistance hysteresis loops measurements of the BL, TI and SRO samples versus magnetic field at low temperature of 1.85$\pm$0.05 K. Fig. 3 depicts such measurement results of the 15 nm BST on 15 nm SRO bilayer, together with the non-hysteretic R versus H curve of a second reference film of the TI (15 nm thick BST on (100) STO wafer) \cite{Koren1}. By increasing the field after ZFC from 0 to 4 T, the resistance shows a peak at 0.26 T which results from the combined effect of the weak anti-localization (WAL) in the 15 nm BST layer where the magnetoresistance increases with field \cite{Koren1}, and the resistance decrease due to realignment of the magnetic domains in the SRO layer of the bilayer. The latter originates in reduction of the overall domain walls resistance \cite{Klein} due to merging of domains with increasing field into a single domain. We also note here that the part of the decrease of the magnetoresistance with increasing field is due to a parabolic background which was clearly observed in our samples also above $\rm T_c$ at about 175 K. With decreasing field down to 0 T, R retraces the former curve down to about 1.5 T, and then bends down following the WAL curve of the reference BST film. Further decrease of the field down to -4 T reveals quite a broad peak in the range of -0.5 to -1.5 T with a maximum at -0.85 T, which can be identified as the coercive field of the bilayer. By sweeping the field up to +4 T and back to zero field, a symmetric result is obtained. \\

\begin{figure} \hspace{-20mm}
\includegraphics[height=6cm,width=8cm]{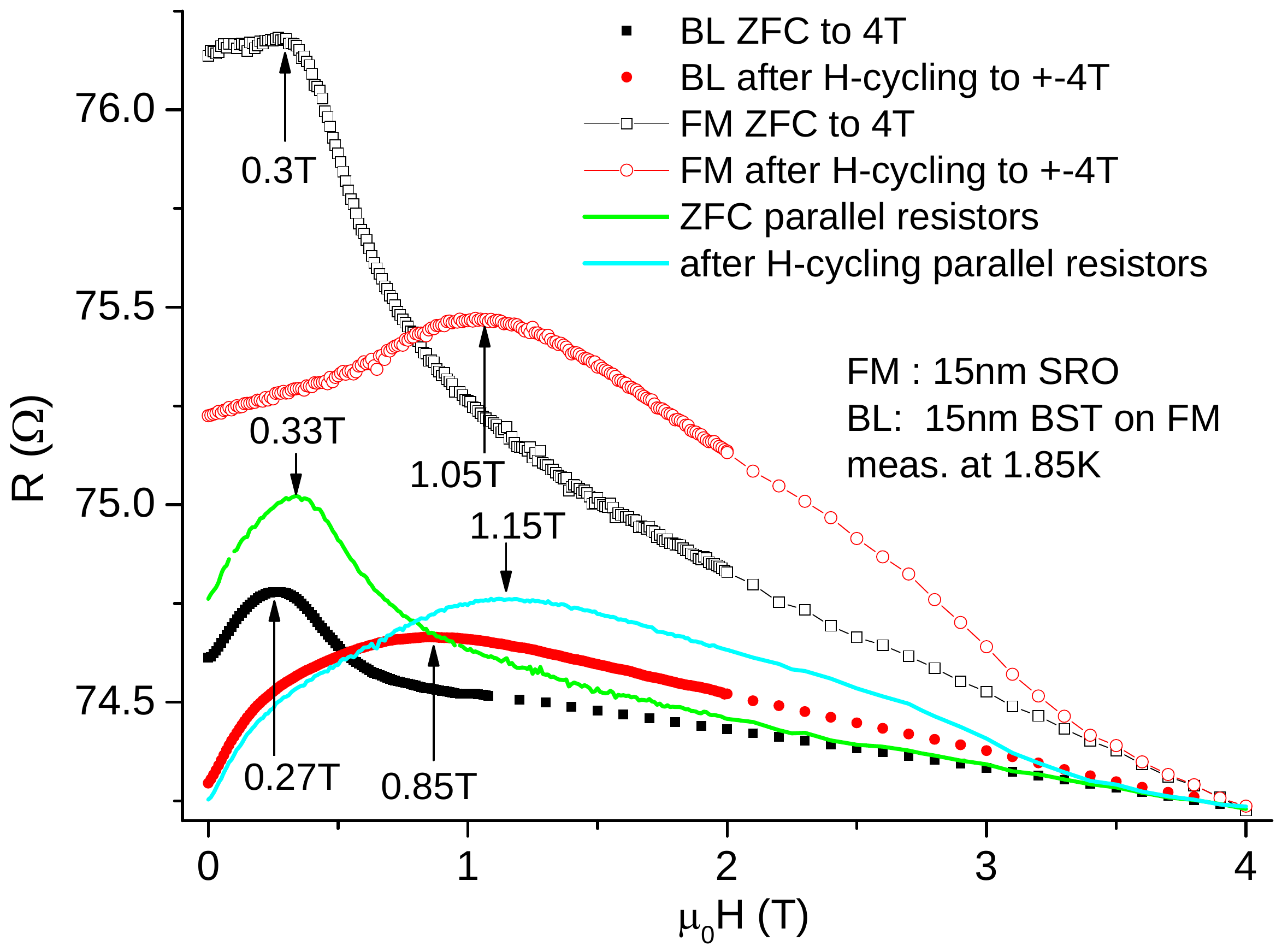}
\hspace{-20mm} \caption{\label{fig:epsart} (Color online) Resistance versus field of the 15 nm BST on 15 nm SRO bilayer under ZFC and after field cycling to $\pm$4 T, together with the corresponding R versus H of the reference 15 nm thick SRO film normalized to the bilayer data at 4 T.  Also shown is the resistance versus field of an hypothetical bilayer of 15 nm BST on 15 nm SRO calculated for two independent resistors connected in parallel with no interaction between the layers. The normalization factors for the SRO and the parallel resistors BST/SRO bilayer are 0.382 and 0.664, respectively.   }
\end{figure}

To check for a signature of MPE in the bilayer, we present in Fig. 4 the positive field data of Fig. 3, together with the R versus H data of the 15 nm thick SRO reference film. All data were normalized at 4 T which is much above the coercive fields of both the BL and SRO samples. One can see that the magnetoresistance after ZFC of the SRO film has only a small increase with field up to a peak at 0.3 T beyond which it goes down with increasing field similar to data in the literature with the measuring bias current parallel to the magnetic domain walls \cite{Klein}. This peak is located at a higher field than that of the bilayer (0.3 vs 0.27 T). A similar shift of the corresponding peak after field cycling (the coercive field peak) is also observed (1.05 T for the SRO film versus 0.85 T for the bilayer). To demonstrate that these shifts are due to MPE in the bilayer, we plotted in Fig. 4 also the expected magnetoresistance of a hypothetical bilayer with no interaction between the layers using a simple addition of parallel resistors of the resistances of the stand-alone SRO and BST films. This yields magnetoresistance peaks at 0.33 T after ZFC and 1.15 T after field cycling. Thus comparison with the magnetoresistance peaks of the real bilayer at 0.27 T and 0.85 T respectively, shows a clear suppression of these coercive fields as expected from MPE in a bilayer where magnetic interactions between the layers are present. \\

\subsection{Meander-lines of 15 nm BST on 40 nm SRO bilayers\\
and 40 nm SRO reference films }

Now we turn to magnetoresistance of bilayers with a thicker SRO base layer. Here the magnetic properties are more robust and similar to those of the bulk SRO, and those of the TI have a much smaller effect on the magnetoresistance results near zero field due to their WAL property. As explained before, we also patterned these bilayers and reference SRO film into long meander-lines in order to increase the sensitivity of the measurements. Fig. 5 shows the magnetoresistance data of a meander-line patterned in one step on a 15 nm thick BST on 40 nm thick SRO bilayer. As one can see the data is quite similar to results on thick SRO films where the measuring bias current is perpendicular to the orientation of the domain walls (no upturn near zero field after ZFC) \cite{Klein}. Again, to check for MPE we compared this data with that of the reference 40 nm thick SRO film (see inset). One can see that the coercive field peak shift of the SRO reference film to higher fields compared to that of the BL is now quite small, but the width of the former is about 30\% higher than that of the latter (0.29 T vs 0.22 T). This is a clear signature of MPE, though we know of no theory that explains this behavior. Possibly, magnetic fluctuations are suppressed more in the bilayer than in the stand-alone SRO film due to pinning of the domain walls at the TI-FM interface. \\

\begin{figure} \hspace{-20mm}
\includegraphics[height=6cm,width=8cm]{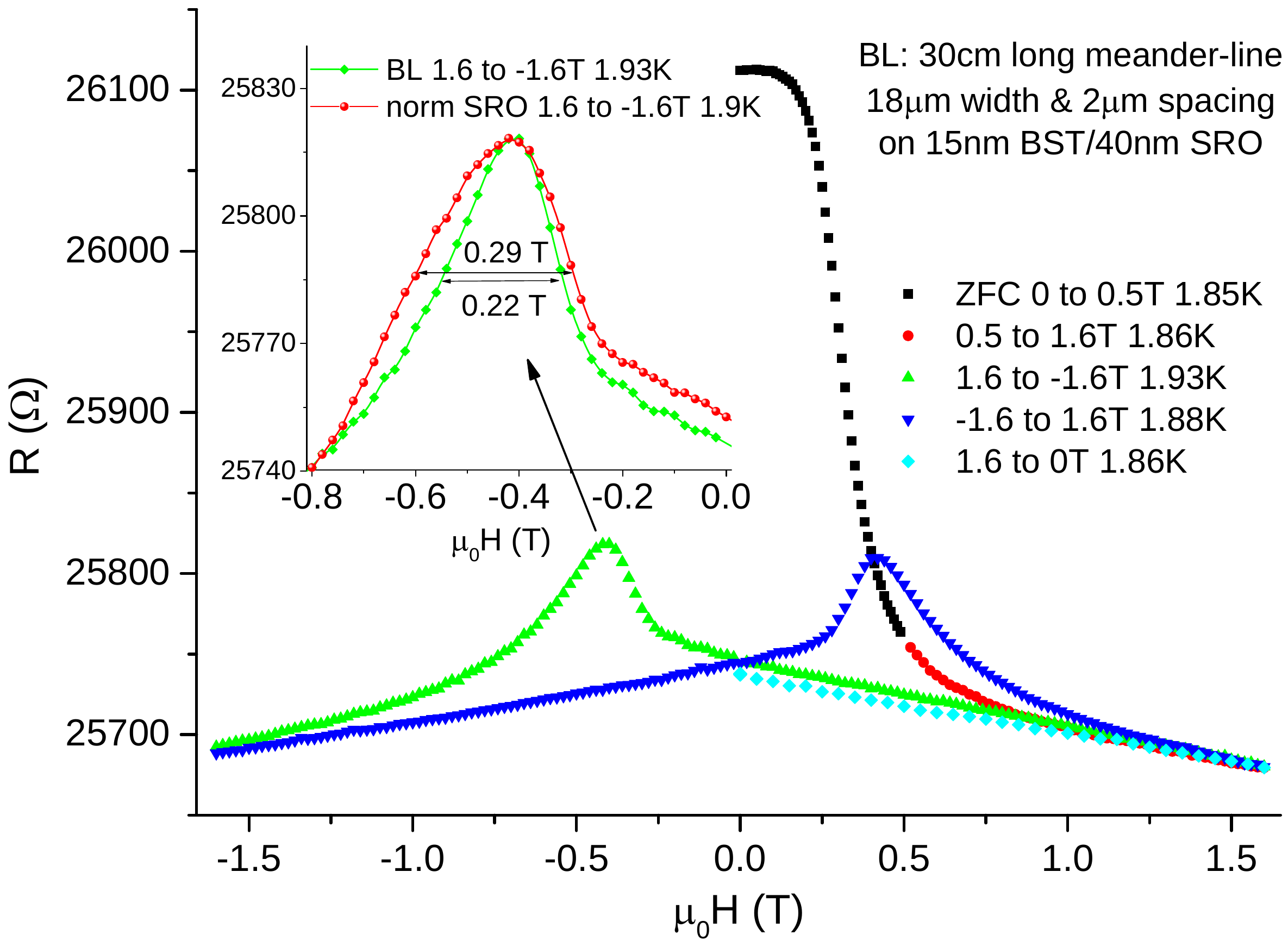}
\hspace{-20mm} \caption{\label{fig:epsart} (Color online) Resistance hysteresis loops versus magnetic field normal to the wafer at 1.89$\pm$0.04 K of a 30 cm long meander-line of 18 $\mu m$ width and 2 $\mu m$ spacing patterned on a 15 nm BST on 40 nm SRO bilayer. The Inset shows a zoom in on the left hand side peak together with the data of the corresponding 40 nm SRO reference meander-line normalized to the maximum peak resistance of the bilayer.  }
\end{figure}

\begin{figure} \hspace{-20mm}
\includegraphics[height=6cm,width=8cm]{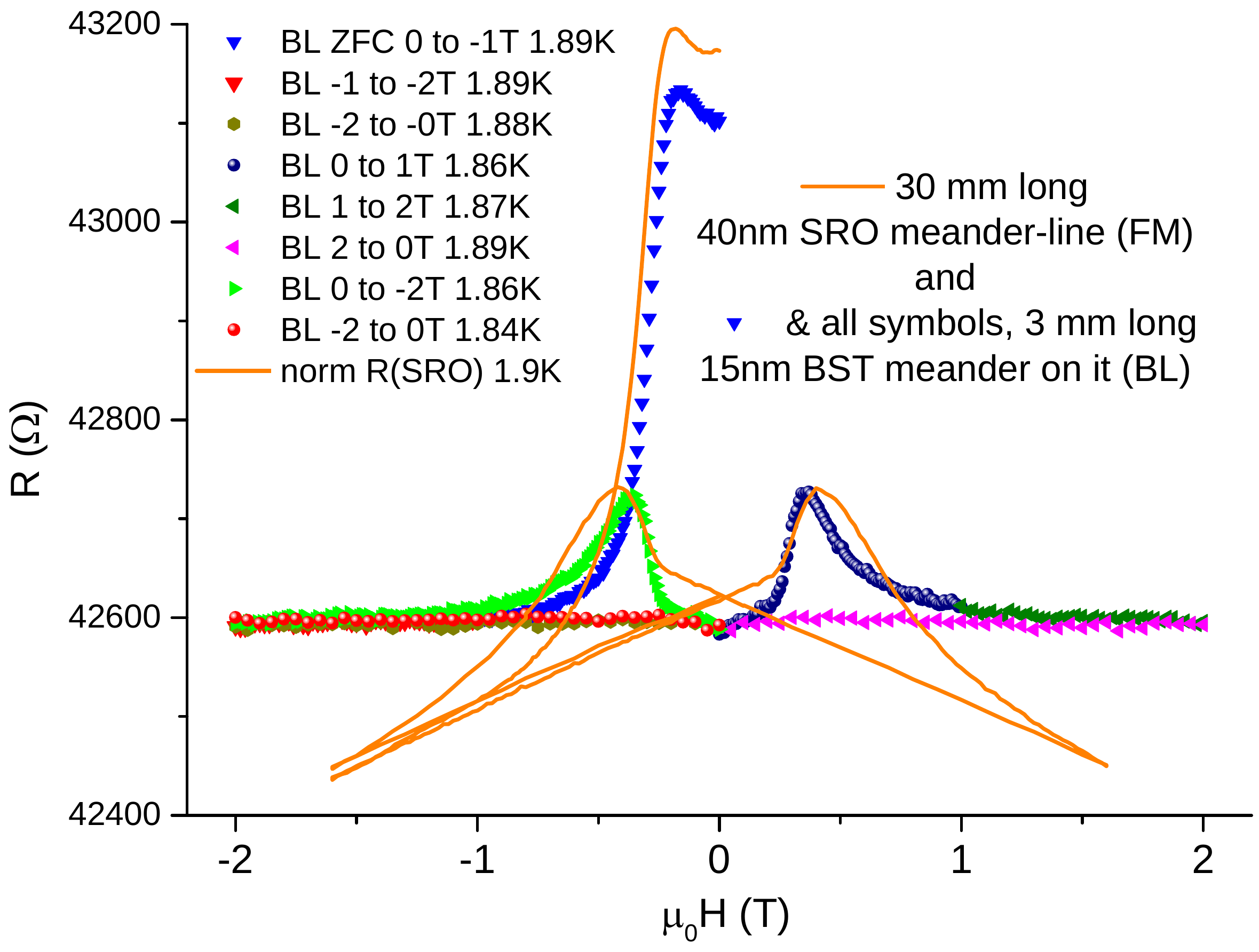}
\hspace{-20mm} \caption{\label{fig:epsart} (Color online) Resistance hysteresis loops versus magnetic field normal to the wafer at 1.87$\pm$0.03 K of two meander-lines of 18 $\mu m$ width and 2 $\mu m$ spacing. One is 30 cm long and patterned on a 40 nm thick SRO layer, and the other is of 15 nm BST deposited on it and re-patterned (the bilayer meander-line). Only a 3 cm long segment of the latter was measured here. The data of the SRO reference meander-line was normalized to that of the peak maxima at the coercive field of the TI-FM bilayer (at $\pm$0.35 T).        }
\end{figure}

Fig. 6 depicts magnetoresistance data of a 15 nm BST on 40 nm SRO bilayer again, but unlike in Fig. 5 the sample was prepared using two separated patterning steps with measurements after each step. First, the base layer of a 40 nm SRO film was deposited and patterned into a 30 cm long meander-line, and its magnetoresistance hysteresis loops were measured. Then the cap layer of 15 nm BST was deposited on it and re-patterned with the same meander-line mask overlapping the first one in the base layer. The magnetoresistance measurements were repeated on the final bilayer, and the data of both measurements at 1.9 K are depicted in Fig. 6. The results of the SRO base meander were normalized to those of the final bilayer at the maxima of coercive field peaks. The shifts and widths of the coercive peaks of the bilayer as compared to those of the SRO meander-line are similar to those of Fig. 5, but both are more pronounced here. The bilayer shifts are larger and the widths are narrower, possibly due to a more resistive interface in Fig. 6. This could originate in residues left over after the patterning process of the base SRO layer, that later affected the results of the bilayer. The flat bilayer background signal in Fig. 6 as compared to Fig. 5 might be a result of this effect. \\

The overall picture that emerges from the results of Figs. 3 to 6 is that the magnetoresistance signal after ZFC is large, its shape versus field can be affected by the TI layer (Figs, 3 and 4), and it apparently masks the smaller peaks at the coercive field $\rm H_{coer}$ which occur at higher fields. We shall thus discuss only the coercive peaks after field cycling. For the bilayers with 15 nm thick SRO we find $\rm H_{coer}$=0.85 T and for the stand-alone SRO layer $\rm H_{coer}$=1.05 T, while for the bilayers with 40 nm thick SRO the corresponding results are 0.41 - 0.36 T and 0.42 - 0.41 T, respectively. Clearly, the $\rm H_{coer}$ of thinner SRO layers are higher than those of the thicker ones, which indicates weaker magnetic interactions and smaller domains size in the thinner films \cite{Klein1}. This affects the $\rm H_{coer}$ values of the bilayers in a similar way. The shifts and widths of the coercive peaks of the bilayers  compared to the reference ferromagnetic films are more interesting. All the magnetoresistance results of the bilayers show shifts of $\rm H_{coer}$ to lower fields and narrowing of the corresponding peaks as compared to the results of the SRO films indicating a clear MPE. Comparing the $\rm H_{coer}$ values of the peak positions in Fig. 4 we find 1.15 T for the non-interacting BL (two resistors in parallel), which is larger than the measured 1.05 T in the SRO reference layer, which is larger still than the 0.85 T of the BL. All these are clear indication that there are strong magnetic interactions between the layers of the bilayer, and demonstrate a strong magnetic proximity effect in the present TI-FM bilayers. \\

\section{Conclusions}

By comparing measurement results of resistance versus temperature and magnetoresistance at low temperatures of TI-FM bilayers to those of a reference SRO film, signatures of magnetic proximity effect in the bilayers were found. These included suppression of the temperature coefficient of resistance (TCR) in the bilayers, as well as shifts of the coercive fields to lower fields, and narrowing of the corresponding coercive peaks in the bilayers as compared to the reference SRO films. The supposedly weak itinerant ferromagnetism of SRO seems to induce a surprisingly robust MPE in BST/SRO bilayers.\\

{\em Acknowledgments:}  We are grateful to Yoichi Ando for useful discussions, and to Yaron Jarach for the EDS measurements of the BST film stoichiometry.\\

\newpage

\noindent\large{\underline{Supplementary Material}}\\
\normalsize

In this supplementary we present structural, morphological and transport properties of a 300 nm thick $\rm Bi_{0.5}Sb_{1.5}Te_3$ film on (0001) Sapphire ($\rm Al_2O_3$), and of the 15 nm $\rm Bi_{0.5}Sb_{1.5}Te_3$ on 15 nm $\rm SrRuO_3$ bilayer on (100) $\rm SrTiO_3$ of the main article. This is done by studying x-ray diffraction data, analyzing Atomic Force Microscopy (AFM) images and by the standard 4-probe dc measurements, respectively. We also present Electron Dispersive Spectroscopy (EDS) results of the laser ablated $\rm (Bi_{x}Sb_{1-x})_2Te_3$ films used in the present study.\\

\section{$\rm Bi_{0.5}Sb_{1.5}Te_3$ film on (0001) $\rm Al_2O_3$}
\normalsize \baselineskip=6mm  \vspace{6mm}

\begin{figure} \hspace{-20mm}
\renewcommand{\figurename}{Fig. S\hspace{-1.5mm}}
\includegraphics[height=9cm,width=14cm]{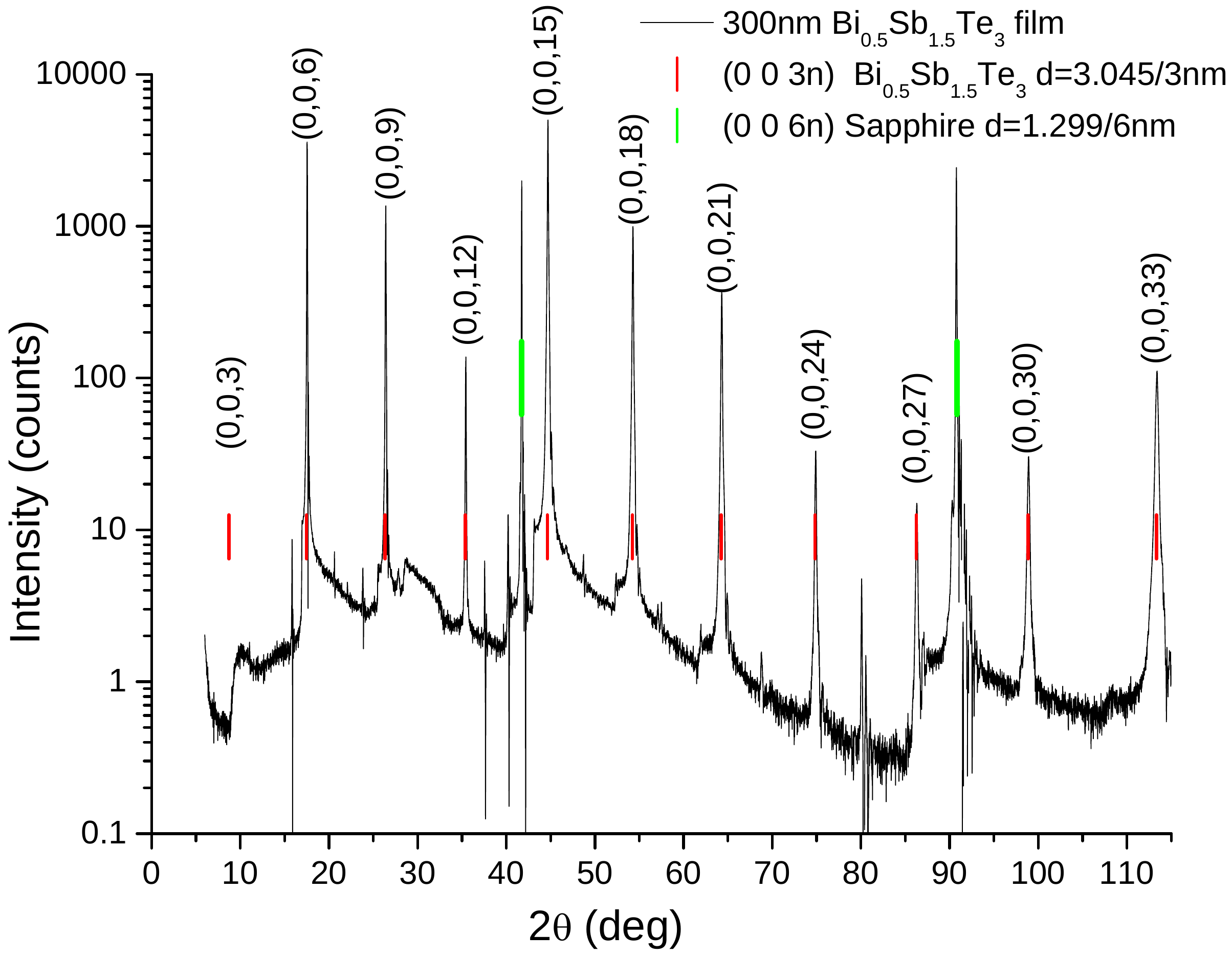}
\hspace{-20mm} \caption{\label{fig:epsart} (color online) X-ray diffraction data of a 300 nm thick BST film on c-axis oriented Sapphire wafer where the scattering intensity versus $2\theta$ is measured in the $\theta - 2\theta$ configuration. The calculated (0,0,3n) peak angles for c=3.045 nm are marked by the vertical red bars, and each peak identity is marked above it. The two (0,0,6n) peaks of the Sapphire substrate are marked by the longer green bars.  }
\end{figure}

To demonstrate the quality of our laser ablated topological films we start with  a 300 nm thick  $\rm Bi_{0.5}Sb_{1.5}Te_3$  (BST) film deposited on c-axis (0001) $\rm Al_2O_3$ (Sapphire). This film was prepared under the same deposition conditions as the BST films in the main article. X-ray diffraction result of this film is plotted in Fig. S7 which shows clearly the (0,0,3n) peaks of the quintupled hexagonal structure (R -3 m :H) of the BST film up to n=11. The measured c-axis lattice constant of this thick film is c=3.045 nm which is slightly smaller than the c = 3.049 nm and 3.046 nm  bulk values of $\rm Bi_2Te_3$ and $\rm Sb_2Te_3$, respectively \cite{Sb2Te3}. The Sapphire substrate (0,0,6n) peaks are also seen and marked in Fig. S7 (with d-spacing of 1.299/6 nm). We thus conclude from the data of this figure that the present BST film is preferentially oriented with c-axis normal to the wafer.\\

\newpage

\begin{figure} \hspace{-20mm}
\renewcommand{\figurename}{Fig. S\hspace{-1.5mm}}
\includegraphics[height=9cm,width=12cm]{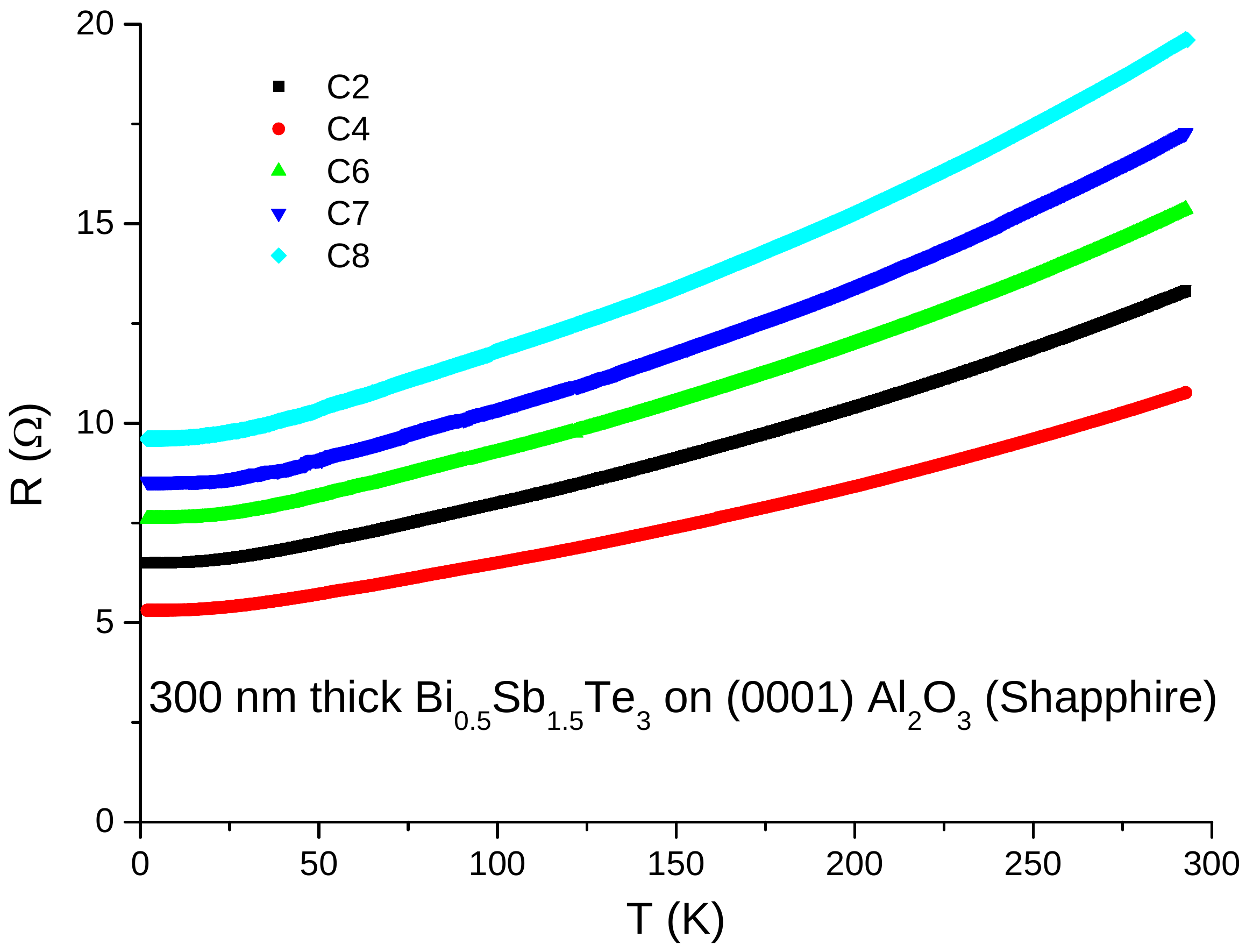}
\hspace{-20mm} \caption{\label{fig:epsart} (Color online) Resistance versus temperature of the 300 nm thick BST film of Fig. S7 at five different locations on the the $\rm 10\times 10\,\,mm^2$ wafer. The contacts are spaced by 1 mm distance between one another along the line separating the wafer into two halves. Results of five of the ten contacts (Ci) are shown here. }
\end{figure}

Fig. S8 shows the resistance versus temperature results of the BST film of Fig. S7. The five different curves are taken at five different location on the $\rm 10\times 10\,\,mm^2$ wafer across the line separating it into two halves. The different resistance of these five contacts is due to their different lead lengths. One can see that the behavior is metallic with a tendency to constant resistance at low temperatures. The metallic behavior is apparently due to Te and Sb vacancies and the saturation at low temperature seems to originate in localization of the charge carriers. There is no upturn of the resistance at low T here as often seen in $\rm Bi_2Se_3$ \cite{Butch}. The corresponding resistivity values of this film range between 0.3 to 0.6 m$\Omega$ cm, which is in the carrier density range of about 10$^{18}$ cm$^{-3}$ when compared to that of $\rm Bi_2Se_3$ \cite{Butch}.\\

\newpage

\section{$\rm Bi_{0.5}Sb_{1.5}Te_3$ on $\rm SrRuO_3$ bilayer on (100) $\rm SrTiO_3$ }
\normalsize \baselineskip=6mm  \vspace{6mm}

\begin{figure} \hspace{-20mm}
\renewcommand{\figurename}{Fig. S\hspace{-1.5mm}}
\includegraphics[height=9cm,width=14cm]{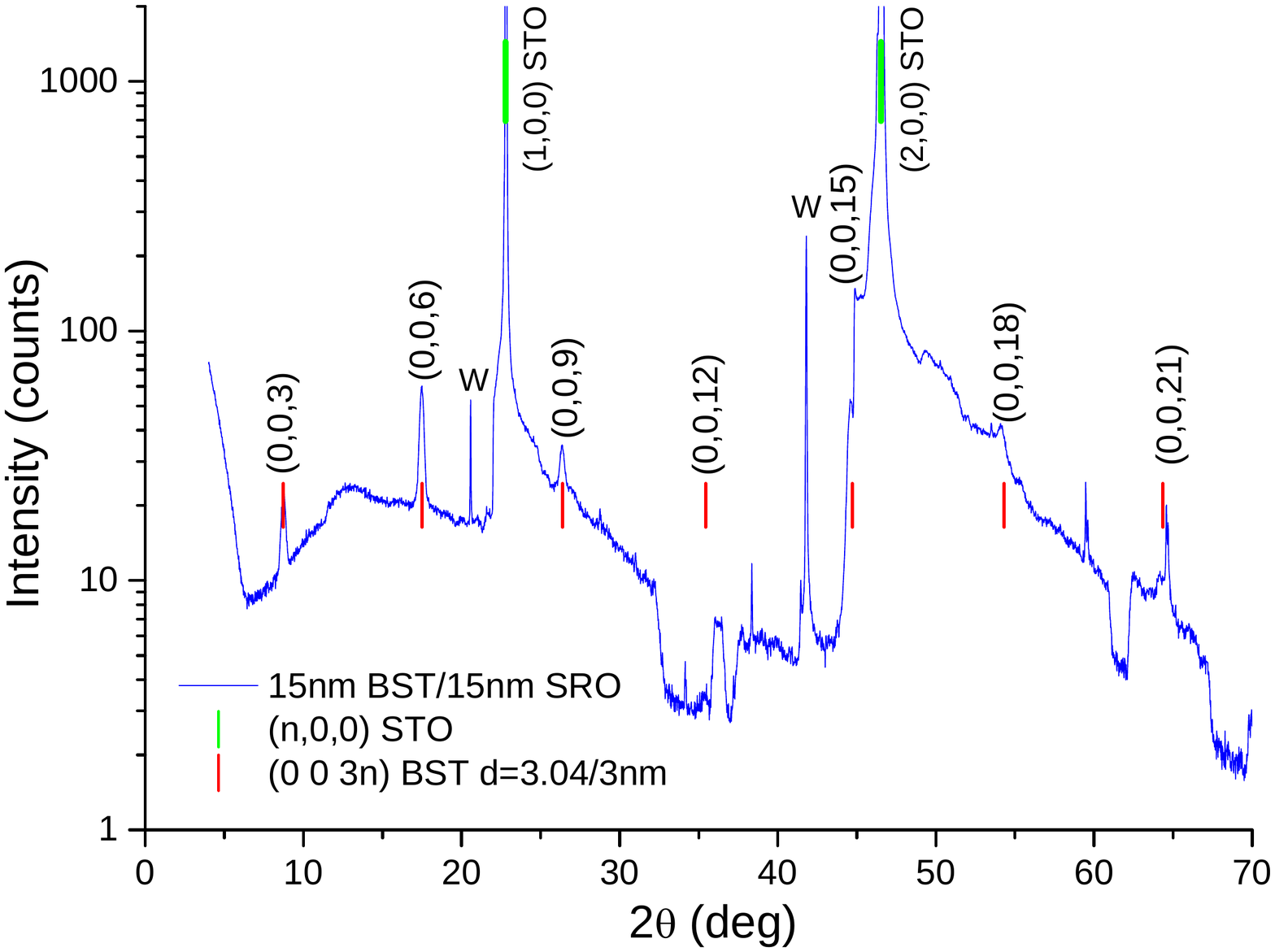}
\hspace{-20mm} \caption{\label{fig:epsart} (Color online) X-ray diffraction results of the bilayer of Figs. 1-4 of the main article. This bilayer is 15 nm BST on 15 nm SRO on (100) STO. The calculated (0,0,3n) peak angles for c=3.04 nm are marked by the vertical red bars, and each peak identity is marked above it. The two (n,0,0) peaks of the (100) STO substrate are marked by the longer green bars.      }
\end{figure}

We turn now to the x-ray diffraction result of the 30 nm thick bilayer of of the main article.  This bilayer is the 15 nm thick BST on 15 nm thick $\rm SrRuO_3$ (SRO) on (100) $\rm SrTiO_3$ (STO) wafer of Figs. 1-4. Fig. S9 present the X-ray diffraction results of this bilayer where again one can see (0,0,3n) peaks of the BST layer. These peaks are quite weak due to the thinness of the BST layer, a mere 15 nm thickness. Since the SRO peaks overlap those of the strong (n,0,0) STO peaks, they could not be resolved \cite{Marshall}. The very narrow W peaks are satellites of the (n,0,0) STO peaks and originate in the W base copper cathode of the x-ray diffractometer. Here again the vertical red bars mark the calculated angles of the (0,0,3n) peaks of BST with c=3.04 nm, and the longer green bars mark the angles of the (n,0,0) STO peaks with a=0.3904 nm. One concludes from the results of Fig. S9 that the BST film is preferentially oriented with the c-axis normal to the wafer.\\

\begin{figure} \hspace{-20mm}
\renewcommand{\figurename}{Fig. S\hspace{-1.5mm}}
\includegraphics[height=14cm,width=10cm]{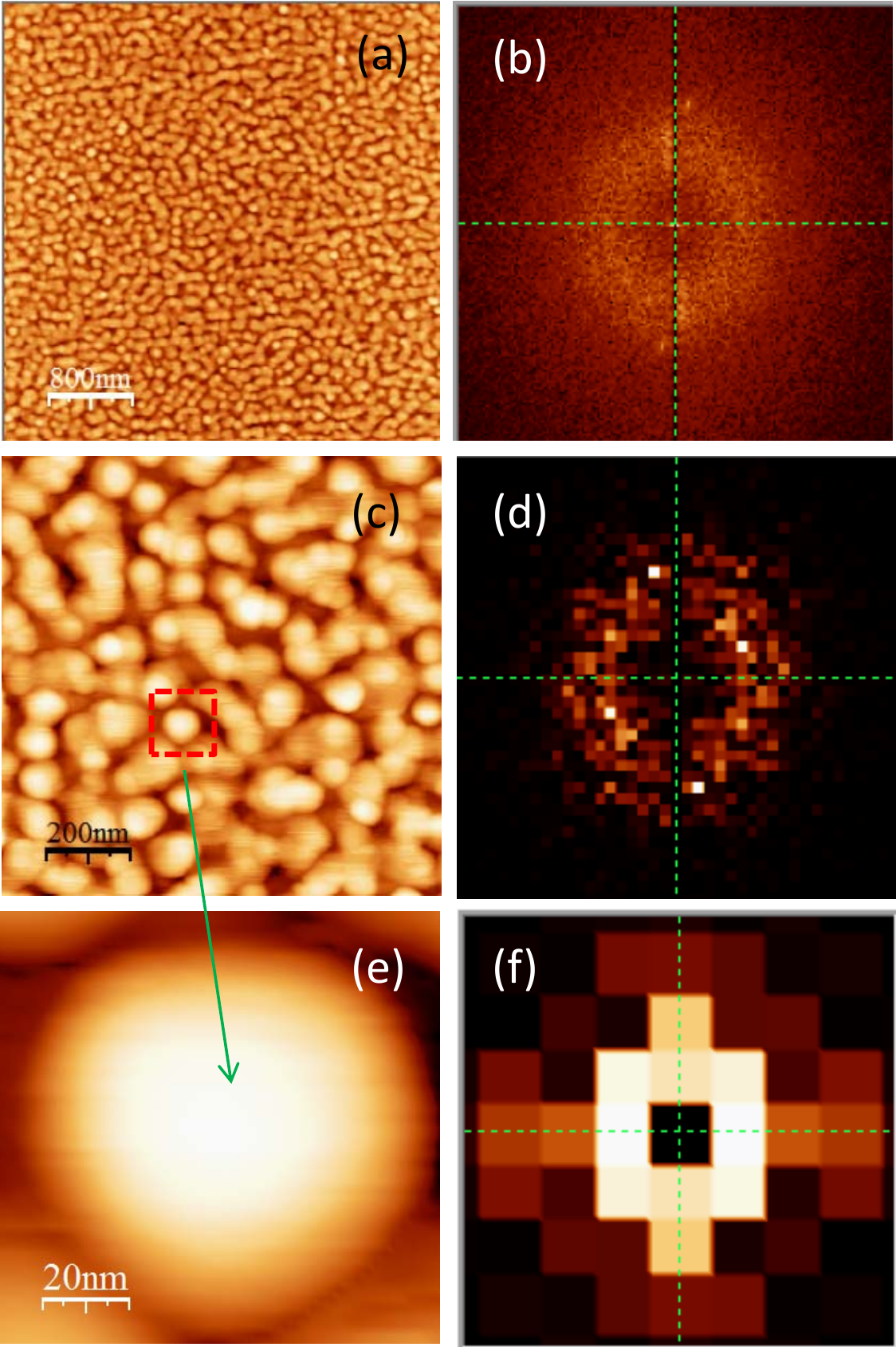}
\hspace{-20mm} \caption{\label{fig:epsart} (Color online)  AFM images of different size areas of the bilayer of Fig. S9 (left side panels), together with their corresponding FFT images (right side panels). Each FFT k-space image is placed side by side with the corresponding real space image. The morphology here is obviously that of the top BST layer of the bilayer.   }
\end{figure}

To study the in-plane orientation of the grains of the BST on SRO bilayer of Fig. S9, we used atomic force microscope (AFM) images. These are presented in Fig. S10 (a), (c) and (e) where the topography is seen in images of different size areas, together with the corresponding Fast Fourier Transform (FFT) images  in (b), (d) and (f), respectively. Fig. S10 (a) and (c) show good crystallization of the top BST layer with root mean square (rms) roughness of 3.8 nm. Compared to the overall 30 nm thickness of the bilayer, this is about $\pm$6 \% roughness and therefore the surface morphology can be considered quite smooth (keep in mind that this is a laser ablated bilayer and not an atomically smooth molecular beam epitaxy (MBE) layer). The FFT images in k-space indicate the in-plane order of the crystallites of the bilayer in real space. Fig. S10 (b) shows an almost isotropic intensity distribution which means random crystallites orientation in the plane of the bilayer of Fig. S10 (a) ("mosaic" in-plane structure). The smaller area scanned in Fig. S10 (c) yields a less isotropic crystallites distribution as can be seen in Fig. S10 (d) where a weak signature of long range hexagonal order can be observed (the six bright spots forming an approximate hexagon). Zoom-in on a single crystallite of Fig. S10 (c) is given in (e). The FFT image of this crystallite is depicted in (f) where a clear hexagonal symmetry is observed. We therefore conclude that on the scale of small crystallites, the long range hexagonal order in the BST top layer is preserved. This however is not the case for larger areas of the bilayer where a mosaic in-plane structure of the crystallites is formed. We also note in this context that the identification of the peaks in the x-ray results of Figs. S7 and S9 was obtained using the hexagonal phase c-axis lattice constant. Thus we conclude that our BST layers are hexagonal, with c-axis preferential orientation, and mosaic in-plane order. \\

\section{Electron Dispersive Spectroscopy (EDS) data}
\normalsize \baselineskip=6mm  \vspace{6mm}

As described in the main article, a $\rm Bi_{0.2}Sb_{1.8}Te_3$ pressed target was used for the deposition of our $\rm (Bi_{x}Sb_{1-x})_2Te_3$ films. For the EDS measurements we used a 300 nm thick film which was cleaved immediately before insertion into the measuring system. Measurements of three different spots on the cleaved area yielded mean atomic percent concentrations ratio of of Bi : Sb : Te = 10.2 : 30.8 : 59.0, with standard deviations of 0.53, 0.77, 0.38, respectively. Within the accuracy of the EDS measurements, this ratio is very close to 10 : 30 : 60 which yields the formula $\rm Bi_{0.5}Sb_{1.5}Te_3$. This chemical composition is abbreviated in the article as BST. The fact that the raw EDS data does not yield this exact concentration ratio (0.5:1.5:3), is apparently due to the presence of Sb and Te vacancies in our films. This is a direct result of the fact that at the low deposition pressure of our films (vacuum of about 5-10$\rm \times 10^{-7}$ Torr), Sb has the highest vapor pressure, Te an intermediate vapor pressure, and Bi the lowest one. \\

\bibliography{AndDepBib.bib}

\bibliography{apssamp}

\end{document}